Paper ID EU-TP1757

# Trinational Automated Mobility

**Jonas Vogt[1*], Niclas Wolniak[1], Prof. Dr.-Ing. Horst Wieker[1]**

1. htw saar, University of Applied Sciences, Saarbrücken, Germany, jonas.vogt@htwsaar.de

**Abstract**

Safe, environmentally conscious and flexible – these are the central requirements for the future mobility. In the European border region between Germany, France and Luxembourg, mobility in the world of work and pleasure is a decisive factor. It must be simple, affordable and available to all. The automation and intelligent connection of road traffic plays an important role in this. Due to the distributed settlement structure with many small towns and village and a few central hot spots, a fully available public transport is very complex and expensive and only a few bus and train lines exist. In this context, the tri-national research project TERMINAL (Automated electric minibuses in cross-border commuter traffic) aims to establish a cross-border automated minibus in regular traffic and to explore the user acceptance for commuter traffic. Additionally, mobility-on-demand services are tested, and both will be embedded within the existing public transport infrastructure.

**Introduction**

Going from and to work in a dense metropolitan area such as Paris or Berlin is easy. A variety of public and private transport possibilities such as buses, trains, trams, sharing vehicles, private vehicles and more are available. Looking at rural areas such as the border region of Germany, France and Luxembourg the situation is more difficult. Some public transport lines with busses and trains exist. However, most commuters use their own vehicle. This poses a problem for people without a driver license or for people who are not able to drive. The first one is especially true for apprentices and trainees who are not old enough to obtain a driver's license or who simple do not see the necessity to have one. The number of people in possession of a driver licenses in the age range from 18 to 29 is slowly but constantly decreasing. [1, p. 23]. Additionally, the number of commuters in the region is increasing. Figure 1 shows the 10 years mean for the numbers of commuters between Luxembourg, the German state Saarland and the France region Lorraine [2]. The triangles indicate the tendency for the last 10 years where an upwards pointing triangle (green) means more commuters, sideways (blue) means the same and a downwards pointing triangle means less commuters. The overall mean is around 100,000 commuters per day with up to 114,000 commuters in 2017 coming from around 91,000 commuters in 2008.

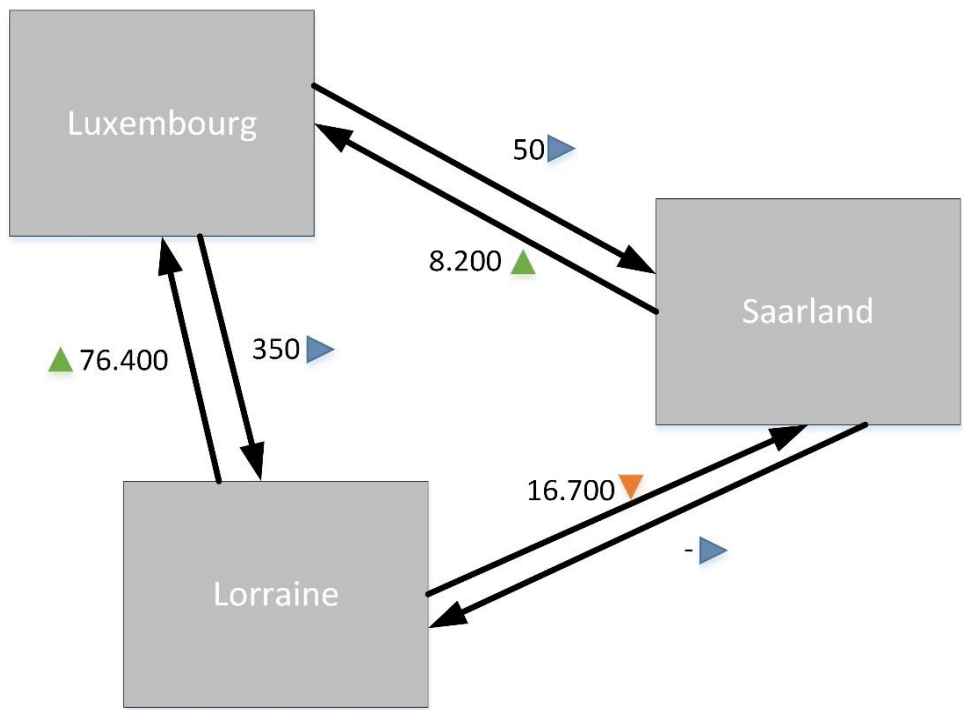

**Figure 1 – Mean number of commuters in the Greater Region over the last 10 years**

In the same time the number of motorways, and bus or train lines did not change significantly. Those numbers indicate a more and more dens traffic on the motorways. Additionally, Luxembourg has with 662 vehicles per 1000 citizens the second largest vehicle numbers per citizens in Europe after Lichtenstein. In comparison, Germany has 555 and France 479 vehicles per 1000 citizens [3]. There are different concepts introduced to get people from using their own vehicle and use public transport. E.g. in December 2018, Luxembourg announced that it would make public transport free of charge for all starting 2020 [4, p. 42-43]. Another approach is to use automated mini buses and mobility on demand services, which represents the idea of the project TERMINAL.

Two application scenarios that extend along the axes Creutzwald-Überherrn and Thionville-Luxembourg were identified. Both are characterised by a significant number of cross-border commuters. Several large industrial and service areas (Häsfeld, etc.) are located around the community of Überherrn, where many workers from France, which is close to the border, also commute. The Thionville-Luxembourg-City axis is also one of the busiest commuter routes in the region.

The objectives for both axes are complementary. At first, data will be aggregated within the French-Luxembourg border area, which will be used to prepare this route for automated traffic and creates additional input for the (partially) automated shuttle, which will operate in the second phase of the project on the Creutzwald-Überherrn route.

Based on the knowledge and the experience gained from these two applications, the project will evaluate the transferability of the tested mobility concepts to other routes and areas in the Greater Region.

Therefore, the project makes it possible to open the emerging innovations in the field of mobility in the decade 2020-2030 in the context of cross-border transport of the Greater Region. It thus makes a fundamental contribution to the sustainable further development of the mobility of cross-border commuters and in particular of trainees (without a driving licence) in the Greater Region and is one of the first concretisations of the German-French-Luxembourg test field on automated and connected driving. Additionally, the project will be a forerunner in automated shuttle services in an international context as there are no other projects cornering this subject, from our current point of view.

In the change of mobility, automatic driving offers some advantages over conventional manual driving, which includes improving road safety to prevent road accidents and improving individual mobility, which causes using mobility instead of owning it. Additionally, a reduction of fuel consumption and pollutant emissions by an efficient use of mobility as a resource.

The project aims for a cross-border automated bus line Überherren Creutzwald, guidance for the introduction of cross-border automated public transport connections for transport companies and provide recommendations for policy action. Those provide cities, regions and the state with information for guidance in mobility policy for cross-border worker mobility.

**Approach**

At first, the framework conditions for the use of automated and connected vehicles, in particular minibus shuttles will be established with a special focus on national differences and the resulting requirements for pilot tests. The participation of approval-relevant bodies (like MWEAV Saarland, DREAL Grand-Est) makes it possible to obtain necessary exemptions for the test operation. The technical and legal aspects also include questions of IT security and data protection, which must be compared against the background of an international deployment scenario and mapped at a legally compliant level for all participants.

Based on this information the infrastructural prerequisites will be established and the vehicle adapted for the specific purpose. This also includes the definition of a secure and reliable communication architecture between the vehicles, the environment (e.g. traffic light) and a cloud solution that collects data from vehicles for further analysis. The decision on the vehicle for the Germany-France test track is made based on the models available at that time, as the market is currently very dynamic. The corresponding vehicle is then prepared for use on the selected test track.

For the France-Luxembourg application, a conventional bus shuttle is equipped with sensor technology so that the relevant data can be obtained for the simulation of an automated application. The legal and technical requirements lay the foundations on which the pilot deployment will be carried out and monitored operationally. Potential users (commuters) and cooperating companies (employers) will be addressed and provided with information in order to make them more aware of the experience of the pilot.

Additionally, an evaluation of the test operation regarding the acceptance and satisfaction of the users will take place. In parallel to the user survey, the collected technical data is evaluated in order to carry out a technical-functional evaluation of the test operation (availability, safety, incidents, etc.).

In order to estimate the potential of the designed and tested mobility services for the entire Greater Region, a transferability analysis will be done. This involves a holistic consideration of the new, automated forms of mobility in interaction with established public and private transport. Both the potentials of scheduled use and on-demand mobility are examined against the background of the current and future socio-demographic situation in the Greater Region, in particular its commuter flows. The experience gained from the project provides a better estimation of the perspective costs associated with the development of such an offer including the effect on the labour demand (drivers).

Due to its innovative dimension, the project can have a significant impact on the perception of automated driving in the public as well as among political decision-makers. Therefore, both groups – in addition to a web-based presentation of the project goals, actions and results – must be addressed and integrated through specific workshops and information events. These will be carried out at the beginning as well as at the end of the trial to discuss the findings and the transfer of the project in the participating sub-regions of the Greater Region.

Finally, it can be stated that the cross-border benefit lies in the prototypical implementation of an innovative transport service for cross-border commuters. Based on the national developments, the project concentrates on the issues arising from cross-border deployment and can thus serve as a blueprint for similar projects in other European border areas. The cross-border dimension is also reflected by the project partners, who contribute their complementary competences to the project.

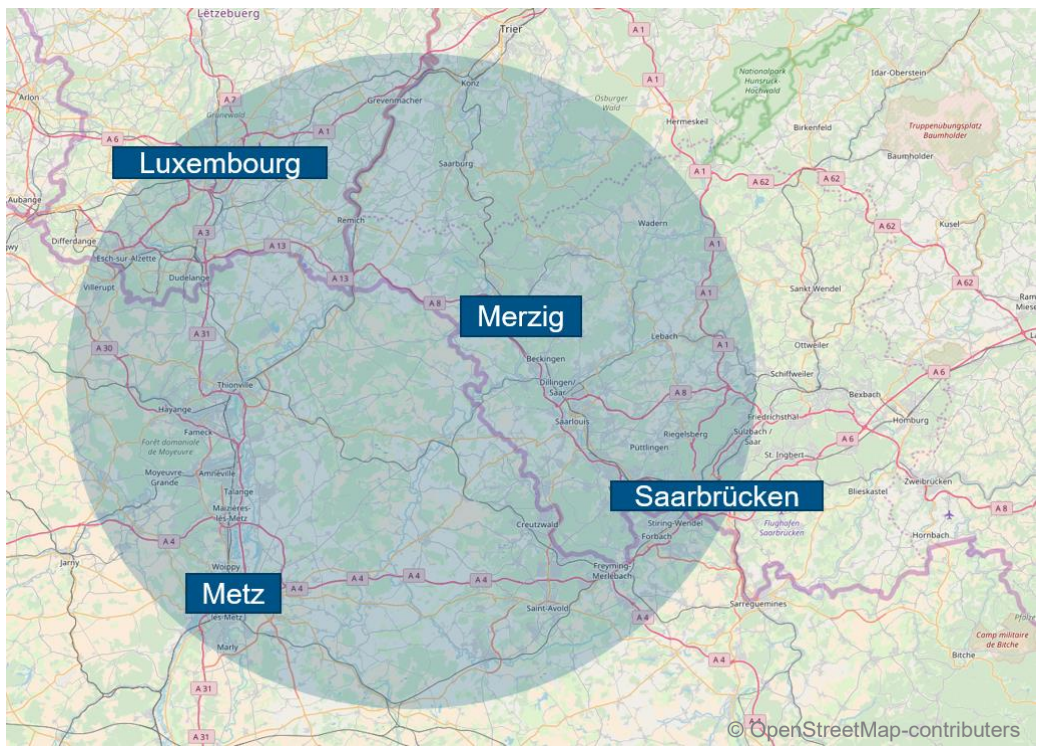

**Figure 2 – Area of the German-French-Luxembourg test field**

**Trinational test field**

The project will be one of the first project to be launched at the "digital test field Germany-France-Luxembourg for automated and connected driving", shown in Figure 2. The test field was established by the ministers for traffic from Germany, France and Luxembourg and announced at the IAA in Frankfurt (Germany) in 2017 [5]. Goal of the test field is to support sustainable, future-oriented traffic and economic politics. For that purpose, four focus areas were developed:

- Continuous compatibility of automated driving functions (functional safety in cross-border traffic)
- Linking of automated driving functions with connected driving, including connection to Intelligent Transport Systems (ITS)
- Impact and effects of automated and connected driving
- Challenges related to the generation, processing, storage, dissemination and exploitation of data for automated and connected driving

**Goals**

The TERMINAL project targets cross-border commuters in the Greater Region. Improving their mobility possibilities and create alternatives to their own vehicles in the form of automated solutions for local and individual public transport. The problems of the current solutions are mainly the lack of flexibility of the mobility offer. Existing cross-border mobility services are rare and based on fixed routes. By using automated vehicles, the routes of the vehicles can be adapted according to demand. In this way, employees and trainees in cross-border areas who do not yet have access to mobility services can gain access to mobility and thus receive additional support. However, not only working people will benefit from automated mobility. In the course of demographic change and urbanisation, there are fewer and fewer mobility offers, especially for older people in rural areas, as they are no longer financially viable. Here, automated buses that can flexible contribute to maintain mobility for elderlies and provide arguments against "rural exodus".

Accordingly, the project also addresses mobility service providers. These include local public transport and private mobility providers who deal with solutions in individual transport, such as taxi companies and car sharing providers. The possibility of using automated vehicles presents new challenges and opportunities for mobility providers, which will be investigated and demonstrated in the project. The developments will change these companies sustainably. This applies in particular to the working environment of the employees. In an automated world, the work of bus drivers will change (e.g. as passenger companions). The European Commission has initiated a process with employee representatives and the trade unions to accompany this development from the outset and to shape it in a socially acceptable manner [6]. The handouts developed in the project for mobility providers to introduce automated bus routes will help to ensure that all providers have the same prerequisites and opportunities to participate in the market for automated providers. At the same time, this will also ensure transferability.

The project is also intended to address the working population of the affected countries Luxembourg, France, and Germany. The aim is to increase public acceptance of public transport in general and automated public transport vehicles in particular. It is also important to address possible fears and resentments towards automated vehicles. The latter is to be achieved through a systematic approach to the topic: first by clarifying the content and then by concrete experience. The municipalities, countries and states are also among the addressees of the project. Through new forms of mobility, public bodies such as the Grand-Est region or the community of Überherrn will be shown further possibilities for political action in the form of recommendations for action. This will provide the political actors with the information they need to influence mobility, work and the places where people live in the respective regions. Politicians will also be provided with the information to be able to drive the topic of mobility forward as a topic of the future. These findings can also be transferred to other regions.

The last target group is business and research in the Greater Region. The project is intended to promote cooperation in the field of research beyond the scope of the project. The project is also intended to provide impetus for the local economy.

**Conclusion**

The project results form the basis for the future planning of automated public transport services with a special focus on cross-border areas. This involves technical application possibilities, the choice of suitable operating models as well as an estimation of demand potential. These questions will arise in many regions with increasing vehicle numbers, advancing demographic change and increasing cost pressure on public transport service providers. It can be assumed that many new bus models with extended automated capabilities will come into the market in the coming decades. These can then presumably be used independently of the route and thus open new possibilities. Demographic change will lead to a decline in the population, especially in rural regions. As a result, in many places, only a care of existence is possible, and these areas are becoming increasingly unattractive.

In addition, public transport providers are always confronted with rising costs, especially for employees. Due to their presumably more favourable operation, automated transport services can be used to expand or increase the flexibility of services on the one hand by reducing employee's costs and on the other hand to serve previously unprofitable routes for the first time. This is the case above all in border regions, since in many places the demand for cross-border transport is too small or too dispersed, which sometimes constitutes an obstacle to taking up employment in the neighbouring country. Both variants can lead to an improved public transport offer and, under certain circumstances, will lead to a decline in motorised private transport in the region under consideration and open new employment opportunities for commuters.

The results of the project will allow planners to design automated bus services for any area in the future. On the one hand, it provides answers on the technical design of the vehicles and the infrastructure and on the other hand, suitable operating models can be found, and demand estimates made. Through the interaction of a practical test with an associated user survey on one side and modelling on the other, a tool is being developed that will make it possible to examine future possibilities for the implementation of automated public transport services. In this respect, the project results make a valuable long-term contribution.

Through the implementation of a cross-border automated bus line during the project period in combination with the results of the parallel surveys and evaluations, important insights and experiences will be gained. These will result in handouts for transport companies and recommendations for action for politicians.

The responsible political authorities (ministries, municipalities, counties, departments) involved in the project can take measures based on the recommendations for action in order to improve the framework conditions for automated transport and, if necessary, intervene in a controlling manner in developments. Both documents, the recommendations for action and the handouts, will be prepared in such a way that the findings from the project can also be transferred to other regions.

The changes in the work world resulting from automated mobility will be discussed with the trade unions. The employees of the participating countries will be part of those changes so that the changes can be made socially acceptable.

The test operation will give citizens, especially commuters using the service, the opportunity to gain experience and get used to the still new technology of automated driving. This is particularly important as scepticism arises with new technologies [7] and is also pushed due to some fatal accidents in the recent past [8].

**Acknowledgement**

The project TERMINAL is supported by INTERREG V A Greater Region with funds of the European Regional Development Fund (ERDF). The TERMINAL consortia consist of University of Luxembourg, University Lorraine, Technical University Kaiserslautern, Utopian Future Technologies SA, Ministry for Economy, Labour, Energy and Traffic Saarland, and the University of Applied Sciences Saarbrücken – htw saar (consortia lead).